\documentclass[showpacs,amsmath,nofootinbib,amssymb,epsfig]{revtex4}

\usepackage{graphicx}
\usepackage{dcolumn}
\usepackage{bm}

\newcommand\be{\begin{equation}}
\newcommand\ee{\end{equation}}
\newcommand\bea{\begin{eqnarray}}
\newcommand\eea{\end{eqnarray}}
\newcommand\bseq{\begin{subequations}} 
\newcommand\eseq{\end{subequations}}
\newcommand\bcas{\begin{cases}}
\newcommand\ecas{\end{cases}}
\newcommand{\p}{\partial}

\usepackage{color}

\begin{document}

\title{Einstein static universe from GUP }
\author{K. Atazadeh }\email{atazadeh@azaruniv.ac.ir}
\author{F. Darabi}\email{f.darabi@azaruniv.ac.ir}
\affiliation{Department of Physics, Azarbaijan Shahid Madani University, Tabriz, 53714-161 Iran}

\date{\today}
\begin{abstract}
We consider the existence and stability of the Einstein static universe under the Generalized Uncertainty Principle (GUP) effects.
We show that this solution in the presence of perfect fluid with a minimal length is cyclically stable
around a center equilibrium point. By taking linear homogeneous perturbations, we find that the scale factor of Einstein static universe for closed deformed isotropic and homogeneous FLRW universe depends on the GUP coupling parameter $\alpha$. Thus, in the model by GUP effects, our universe can stay at the Einstein static state past-eternally, which means that the big bang singularity might be resolved successfully by an emergent scenario.

\end{abstract}

\maketitle

\section{Introduction}
Existence of a minimal length below which no other length can be
observed, is a prediction of quantum theory of gravity \cite{a}-\cite{f}. In the framework of the perturbative string theory {\cite{a,b}}, such a minimal observable length is due to the fact that strings cannot probe distances smaller than the
string size. The quantum effects of gravitation at the scale of this minimal length become as important as the electroweak and strong
interactions. Thus, in the framework of high energy physics phenomena such as early universe or  strong gravitational
fields of a black hole, we must consider the drastic effects of such a minimal length.

Deformation of standard Heisenberg commutation relation in the ordinary quantum mechanics is a remarkable feature of the existence of a minimal length {\cite{g,h}}. Such relations are known as the generalized uncertainty principle (GUP). In one dimension,
the simplest form of such relations in the context of the Snyder non-commutative space \cite{ss} can be written as
\begin{equation}\label{uncrel}
\Delta q\Delta p\geq \frac{1}{2}|<\sqrt{1-\alpha p^2}>|,
\end{equation}
which is reduced to  the minimal uncertainty relation when $\alpha<0$. Also, at the first order in $\alpha$, the string theory result  $\Delta q\gtrsim(1/\Delta p+l{_s}^{2}\Delta p)$ \cite{i}, in which the string length $l_{s}$ can be determined with $(-\alpha/2)^{1/2}$, is recovered. Furthermore, if $\alpha>0$ a vanishing uncertainty in the non-commutative coordinate is allowed and appears as soon as $\Delta p$ reaches the critical value of $(\Delta p)^{\ast}=\sqrt{(1-\alpha< p>)/\alpha}$. Thus, the commutation relation can be written as

\begin{equation}\label{xp}
[q,p]=i\sqrt{1-\alpha p^{2}},
\end{equation}
and the only freedom remains on the sign of the deformation parameter $\alpha$. We can then conclude that, a maximum momentum or a minimal length are predicted by the Snyder-deformed relation (\ref{xp}) if $\alpha>0$ or $\alpha<0$, respectively. Various applications of the low energy
effects of the modified Heisenberg uncertainty relations have been
extensively studied, see for example \cite{12}-\cite{16}. Also, in \cite{marco}, the author has considered the implications of a deformed Heisenberg algebra on the Friedmann-Lema\^{\i}tre-Robertson-Walker  cosmological models.

The idea, which uses the Einstein static state to solve the problem of big bang singularity, was first proposed by Ellis et al., and since then it was named the emergent scenario \cite{3,4}. It is easy to see that the existence of a stable Einstein static state universe is a prerequisite for the emergent theory. Otherwise our universe is impossible to stay at the static state past-eternally. The emergent mechanism is unsuccessful for the avoidance of big bang singularity in the theory of general relativity because the Einstein static state solution is unstable. In the very early universe, due to the
fact that the cosmic energy density is very large, it is reasonable to consider some other effects, such as those from quantum gravity and modified gravity, which might help to stabilize the Einstein static state.
Finally,  the stability of the Einstein static state has been studied
in various cases \cite{Carneiro, Mulryne, Parisi, Wu,
Lidsey, Bohmer2007, Seahra, Bohmer2009, Barrow,
Barrow2009, Clifton,Boehmer2010, Boehmer20093, Wu20092,
Odrzywolek, Labrana}, from loop quantum gravity \cite{Mulryne, Parisi, Wu} to $f(R)$ gravity  \cite{Boehmer2010} and $f(T)$ gravity \cite{ft}, from Horava-Lifshitz gravity \cite{Wu20092,Odrzywolek} to brane gravity \cite{Lidsey,ata1} and massive bigravity \cite{mass}. Also, recently  the stability of the Einstein static state has been considered in \cite{ata2,ata3}.

In this paper, we consider the stability of the Einstein static universe in the Friedmann-Lema\^{\i}tre-Robertson-Walker (FLRW) space-time in the
framework of the GUP effects. Section II is devoted to the study of the modified FLRW cosmological dynamics. In Section III we present an analysis of the equilibrium of Einstein solution in the presence of matter. Next, we study a numerical example,
in which the energy contain relativistic matter. In section IV we consider the model under the inhomogeneous perturbations. The paper ends with a brief conclusions in Section V.

\section{Deformed FLRW dynamics}

Here, we first rewire the Snyder-deformed dynamics of the isotropic and homogeneous cosmological models that it has been presented in \cite{marco}. We study the system at classical level searching for the modifications influenced by the deformed Heisenberg algebra. Thus, we consider the ordinary FLRW dynamics and then focus on the deformed one.

\subsection{Standard Friedmann equation}

The isotropic and homogeneous FLRW cosmological models are characterized by the line element
\begin{equation}
ds^{2}=-N^{2}dt^{2}+a^{2}\left(\frac{dr^{2}}{1-Kr^{2}}+r^{2}d\Omega^{2}\right),
\end{equation}
where $a=a(t)$ and $N=N(t)$ are the scale factor and the lapse function, respectively. The scale factor is the only degree of freedom of the system describing the expansion of the universe while the lapse function does not play a dynamical role. The spatial curvature $K$ can be zero or $\pm1$ depending on the topology of the space. The dynamics of such models is encapsulated in the scalar constraint
\begin{equation}\label{scacon}
{\cal H}=-\frac{2\pi G}{3}\frac{p_{a}^{2}}{a}-\frac{3}{8\pi G}aK+a^{3}\rho=0,
\end{equation}
where $G=l_{P}^2$ is the gravitational constant, $\rho=\rho(a)$ denotes for generic energy density of the system and $p_{a}$ is the momentum conjugate to the scale factor $a$. Because of the isotropy, the phase space of general relativity is reduced to 2-dimensional space in which the only non-vanishing Poisson bracket is $\{a,p_{a}\}=1$. By using the Hamilton equations with respect to the extended Hamiltonian, the Friedmann equation can be obtained
\begin{equation}\label{extham}
{\cal H}_{E}=\frac{2\pi G}{3}N\frac{p_a^2}{a}+\frac{3}{8\pi G}NaK-Na^{3}\rho+\lambda\pi,
\end{equation}
where $\lambda$ is a Lagrange multiplier and the last term $\lambda\pi$ is written because the momenta conjugate ($\pi$) to the lapse function vanishes. The equation of motion for $N$ is obtained as  $\dot{N}=\{N,{\cal H}_E\}=\lambda$, and the Hamiltonian constraint (\ref{scacon}) is obtained requiring the constraint $\pi=0$ to be satisfied at all times, i.e. by imposing that the secondary constraint $\dot\pi=\{\pi,{\cal H}_E\}={\cal H}=0$ holds. The other equations of motion with respect to ${\cal H}_E$ read as following
\begin{eqnarray}\label{eqap}
\dot{a}=\{a,{\cal H}_E\}=\frac{4\pi G}{3}N\frac{p_{a}}{a}, \qquad \dot{p_{a}}&=&\{p_{a},{\cal H}_E\}= N\left(\frac{2\pi G}{3}\frac{p_{a}^2}{a^{2}}-\frac{3}{8\pi G}K+3a^{2}\rho+a^{3}\frac{d\rho}{da}\right).
\end{eqnarray}
By using of the above equations and the Hamiltonian constraint (\ref{scacon}), we can obtain the equation of motion for the Friedmann equation as
\begin{equation}\label{canfri}
\left(\frac{\dot{a}}{a}\right)^{2}=\frac{8\pi G}{3}\rho-\frac{K}{a^{2}},
\end{equation}
which is the desired Friedmann equation in a synchronous reference frame, i.e. defined in the $3+1$ framework by $N=1$ and $N^{i}=0$, the time coordinate is identified with the proper time at each point of space. We know that this equation leads to the big-bang singularity where the (general-relativistic) description of the universe is no longer appropriate and quantum modifications are required.

\subsection{Deformed Friedmann equation}

 Now, we are ready to consider the analysis of the deformed dynamics of the FLRW models and specifically we study the one-dimensional case of the scheme considered above. In other words, we check the modifications resulting from the algebra (\ref{xp}) on the classical trajectory of the universe represented in the previous subsection.

The modified symplectic geometry resulting from the classical limit of (\ref{xp}), is the origin of Snyder-deformed classical dynamics and the parameter $\alpha$ is regarded as an independent constant with respect to $\hbar$. According to Dirac's prescription, it is possible to replace the quantum-mechanical commutator (\ref{xp}) via the Poisson bracket
\begin{equation}\label{pm}
-i[\tilde{q},p]\Rightarrow\{\tilde{q},p\}=\sqrt{1-\alpha p^{2}}.
\end{equation}
This relation corresponds exactly to the unique (up to a sign) possible realization of the Snyder space. To obtain the deformed Poisson bracket, some natural requirements must be considered. So, it must possess the same properties as the quantum mechanical commutator, {\it i.e.} it has to be anti-symmetric, bilinear and satisfy the Leibniz rules as well as the Jacobi identity. Thus, the Poisson bracket in the two-dimensional phase space is
\begin{equation}
\{F,G\}=\left(\frac{\p F}{\p \tilde{q}}\frac{\p G}{\p p}-\frac{\p F}{\p p}\frac{\p G}{\p \tilde{q}}\right)\sqrt{1-\alpha p^{2}}.
\end{equation}
Specially, the canonical equations for coordinate and momentum from the deformed Hamiltonian ${\cal H}(\tilde{q},p)$, are given by
\begin{eqnarray}
\dot{\tilde{q}}&=&\{\tilde{q},{\cal H}\}=\frac{\p{\cal H}}{\p p}\sqrt{1-\alpha p^{2}},\nonumber\\ \qquad \dot{p}&=&\{p,{\cal H}\}=-\frac{\p{\cal H}}{\p \tilde{q}}\sqrt{1-\alpha p^{2}}.
\end{eqnarray}
Now, we apply this scheme to the FLRW model in the presence of the matter energy density, namely to the extended Hamiltonian (\ref{extham}). Thus we assume the minisuperspace as Snyder-deformed and consequently, the commutator between the scale factor $a$ and its conjugate momentum $p_{a}$ is uniquely given by
\begin{equation}\label{ap}
\{a,p_{a}\}=\sqrt{1-\alpha p_{a}^{2}}\,,
\end{equation}
 with respect to which the equations of motion $\dot{N}=\{N,{\cal H}_E\}=\lambda$ and $\dot{\pi}=\{\pi,{\cal H}_E\}={\cal H}=0$ are not changed. Indeed, the Poisson bracket $\{N,\pi\}=1$ is not influenced by the deformations induced by the $\alpha$ parameter. Nevertheless, the equations of motion (\ref{eqap}) can be modified in such approach via the relation (\ref{ap}), and we have
\begin{eqnarray}\label{eqapgup}
\dot{a}&=&\{a,{\cal H}_E\}=\frac{4\pi G}{3}N\frac{p_{a}}{a}\sqrt{1-\alpha p_{a}^{2}}, \\\nonumber \qquad \dot{p_{a}}&=&\{p_{a},{\cal H}_E\}=\sqrt{1-\alpha p_{a}^{2}}N\left(\frac{2\pi G}{3}\frac{p_{a}^{2}}{a^{2}}-\frac{3}{8\pi G}K+3a^{2}\rho+a^{3}\frac{d\rho}{da}\right).
\end{eqnarray}
The equation of motion in the canonical case for the Hubble rate can be obtained by solving the constraint (\ref{scacon}) with respect to $p_a$ and then studying the first equation of (\ref{eqapgup}). Explicitly, it has the following form (taking $N=1$)
\begin{equation}\label{deffri}
\left(\frac{\dot{a}}{a}\right)^{2}=\left(\frac{8\pi G}{3}\rho-\frac{K}{a^{2}}\right)\left[1-\frac{3\alpha}{2\pi G}a^{2}\left(a^{2}\rho-\frac{3}{8\pi G}K\right)\right].
\end{equation}
Also, the conservation equations for the matter component is
given by
\begin{equation}\label{ma}
\dot{\rho}+ 3H( 1+ w)\rho =0\,,
\end{equation}
where $w$ is the equation of state parameter of the background matter.
Equation (\ref{deffri}) is deformed Friedmann equation in which it requires the modification originating  from the Snyder-deformed Heisenberg algebra (\ref{ap}). If we consider the flat FLRW universe ($K=0$), the deformed equation (\ref{deffri}) can be written as \cite{marco}
\begin{equation}\label{modfri}
\left(\frac{\dot{a}}{a}\right)^{2}_{K=0}=\frac{8\pi G}{3}\rho\left(1-\text{sgn}\,\alpha\frac{\rho}{\rho_{c}}\right),
\end{equation}
where $\rho_{_{P}}$ is the Planck  energy density and $\rho_{c}=(2\pi G/3|\alpha|)\rho_{_{P}}$ is the critical energy density. Note that in the last step the existence of a fundamental minimal length is assumed . One of the most important consequences of all quantum gravity theories is the existence of a fundamental cut-off length which is related to the Planck cut-off length (for a review see \cite{gar}). Therefore, it is anticipated that the scale factor (the energy density) has a minimum (maximum) at the Planck scale.

The impact of deformed Heisenberg algebra on the Friedmann equation (\ref{modfri})
results in the modifications  manifested in the form of a $\rho^{2}$-term. If $\alpha>0$ and $\rho=\rho_{c}$ in high energy regime,  the Hubble rate vanishes and the Universe experiences a bounce  in the scale factor. Also, the standard Friedmann equation (\ref{canfri}) for $k=0$, is recovered for energy density much smaller than $\rho_{c}$. When  $\alpha=0$, the correction term vanishes and the ordinary behavior of the Hubble parameter is obtained.
The Randall-Sundrum braneworld scenario is also recovered for $\alpha<0$ .

\section{The Einstein static solution and stability}

By using the equations (\ref{deffri}) and (\ref{ma}) for the closed FLRW universe ($K=1$), the Raychadhuri equation can be written as\footnote{We have set units $8\pi G=1$.}
\begin{eqnarray}\label{aa}
\ddot{a}=12 w \alpha  \rho ^2 a^5-36 (w+1) \alpha  \rho  a^3+48 \alpha  \rho  a^3-36 \alpha  a-\frac{1}{2} (w+1) \rho  a+\frac{\rho  a}{3},
\end{eqnarray}
where by solving the equation (\ref{deffri}), the matter energy density $\rho$ as a function of $a$ and $\dot{a}$ is given by
\begin{eqnarray}\label{bb}
\rho=\frac{72 \alpha  a^4+a^2\pm\sqrt{a^4-144 a^8 H^2 \alpha }}{24 a^6 \alpha }.
\end{eqnarray}
The Einstein static solution is described by $\ddot{a} = 0 = \dot{a}$. To begin with, we obtain the conditions
for the existence of this solution. From equations (\ref{aa}) and (\ref{bb}), the scale factor and energy density in this case are given by
\begin{eqnarray}\label{uu}
a^{2}_{_{Es}}=\frac{(1-3w)}{36\alpha(1+3w)},~~~~~~\rho_{_{Es}}=\frac{216\alpha(1+3w)}{(1-3w)^{2}}~~~~~\Longrightarrow~~~~ \rho_{_{Es}}=\frac{6}{(1-3w)a^{2}_{_{Es}}}.
\end{eqnarray}

By considering the solutions (\ref{uu}), the existence condition of an Einstein static universe is reduced to the reality condition for $a_{_{Es}}$ and positivity for $\rho_{_{Es}}$, which for a positive $\alpha$ results in the allowed domain of $w$
\begin{eqnarray}\label{www}
-1/3<w<1/3,
\end{eqnarray}
and for $\alpha<0$ we have
\begin{eqnarray}\label{wwwW}
w<-1/3.
\end{eqnarray}

Now, we are going to study the stability of the critical point for the case of positive sign equation (\ref{bb}). For convenience, we introduce two  variables
\begin{eqnarray}
x_1=a,\quad x_2=\dot{a}.
\end{eqnarray}
It is then easy to obtain the following equations
\begin{eqnarray}
\dot{x}_1=x_2,
\end{eqnarray}
\begin{eqnarray}
\dot{x}_2=\left[-\frac{x_1}{6}+\frac{27}{4}\alpha {x_1}^{3}-\frac{3{x_1}}{2}w\left(\frac{1}{3}+\frac{27}{2}\alpha {x_1}^{2}\right)\right]\rho(x_1,x_2)+12\alpha {x_1}^{5}w\rho(x_1,x_2)^{2}-\frac{9\alpha {x_1}}{2}.
\end{eqnarray}
According to these variables, the fixed point, $x_1=a_{Es},\; x_2=0$ describes the Einstein static solution properly. The stability of the critical
point is determined through the eigenvalue of the coefficient matrix ($J_{ij}=\frac{\partial \dot{x}_{i}}{\partial x_{j}} $)
stemming from linearizing the system explained in details by above two equations near the critical point. Using $\lambda^2$ to obtain the
eigenvalue we have
\begin{eqnarray}\label{ll}
\lambda^2=\frac{189}{\frac{3-63 \alpha  \sqrt{\frac{(w (12 w-7)+1)^2}{(3 w+1)^4 \alpha ^2}}}{(9 w-4) \alpha }-\frac{49 \sqrt{\frac{(w (12 w-7)+1)^2}{(3 w+1)^4 \alpha ^2}}}{(1-4w)^2}-\frac{4}{\alpha -4 w \alpha }-\frac{18}{3 w \alpha +\alpha }}.
\end{eqnarray}
In the case of $\lambda^{2}< 0$ the Einstein static solution has a center
equilibrium point, so it has circular stability, which means that small perturbation from the fixed
point results in the oscillations about that point rather than exponential deviation from it. In
this case, the universe oscillates in the neighborhood of the Einstein static solution
indefinitely.
Here the allowed ranges for $w$ with the requirement $\lambda^{2}< 0$ are obtained in Table 1.
\vspace{5mm}
\begin{center}
{\scriptsize{ Table 1: }}\hspace{-2mm} {\scriptsize Allowed ranges for $w$.}\\
    \begin{tabular}{|l| l |l |  p{800mm} }
    \hline
   {\footnotesize$~~~~~~~~~~~~~~~~~~~~~~~~~~~~~~~~~w$ }& ~~{\footnotesize~ $\alpha$ }  \\ \hline
{\footnotesize ~~~~~~~~~~~~~$-1/3 < w < 1/4~~ {\rm or}~~ w > 1/3$} & ~~{\footnotesize $\alpha>0$} \\\hline
 {\footnotesize ~~~$-1/3 < w < 1/4 ~~ {\rm or}~~  1/4 < w < 1/3 ~~ {\rm or}~~  w > 4/9$} & ~{\footnotesize  ~$\alpha<0$}
\\ \hline
    \end{tabular}
\end{center}

Thus, the stability condition is determined by $\lambda^{2}< 0$ (Table 1). For $\alpha> 0$, this means
that $-1/3 < w < 1/4$. Comparing this inequality with the conditions for existence of the Einstein
static solution (\ref{www}), we find that the Einstein universe is stable for $-1/3 < w < 1/4$.
Especially, it is stable in the presence of ordinary matter ($ w$) and the GUP effects.
Also, for more clarification about the explicit behavior of $\lambda^{2}$ as a function of $w$ and $\alpha$ we have plotted it in Fig.1.

\begin{figure*}[ht]
 \centering
\includegraphics[width=3in]{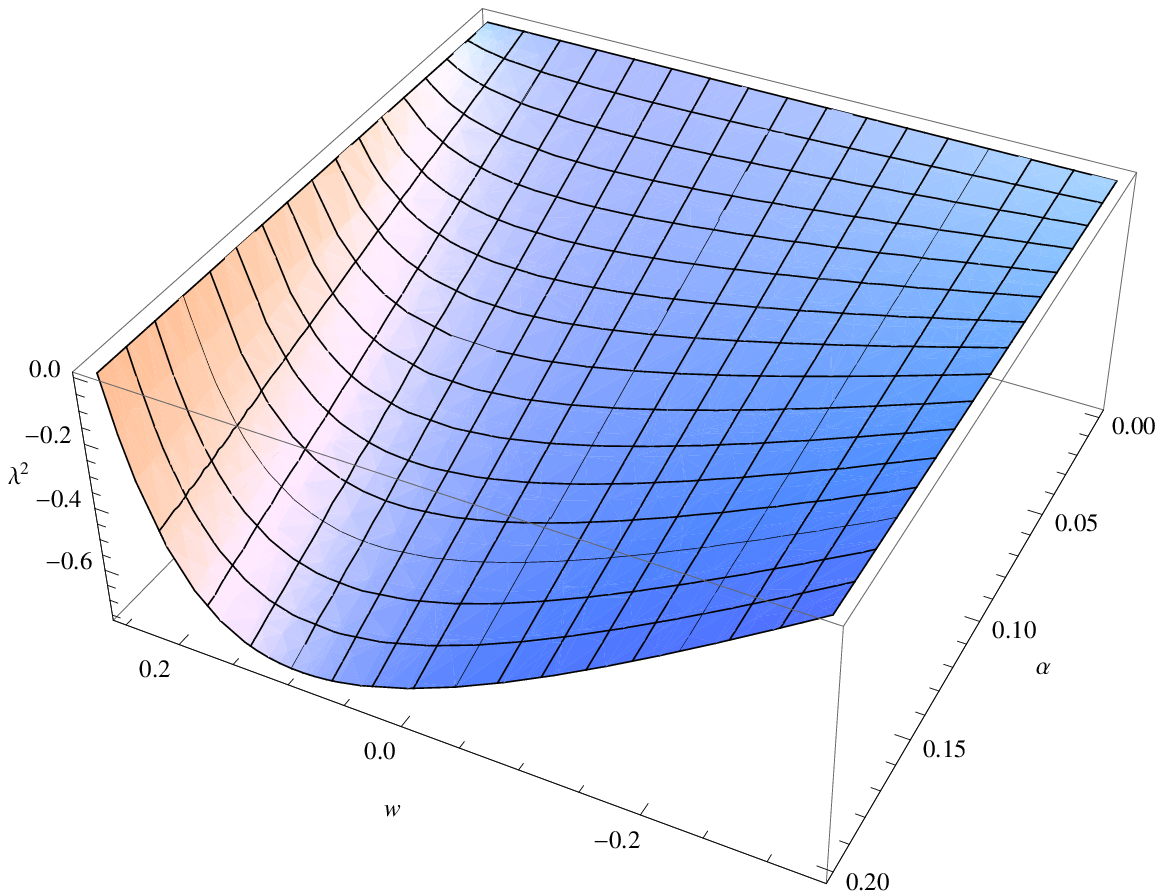}~~~~~~~~~~~~~~~
   \includegraphics[width=3in]{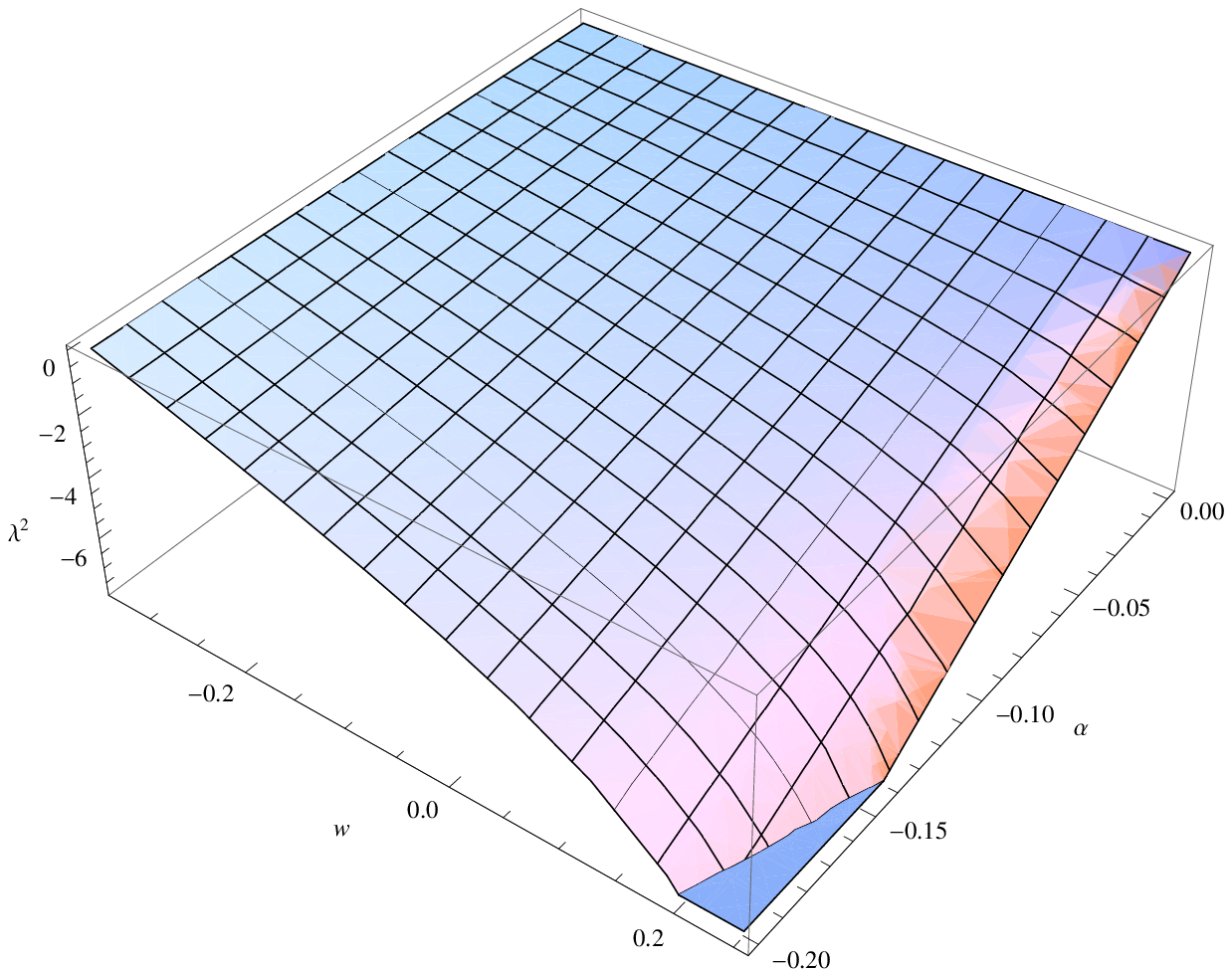}
  \caption{The behavior of $\lambda^{2}$ as a function of $w$ and $\alpha$ for $-1/3<w<1/4$, $0<\alpha<0.2$ (left) and  $-0.2<\alpha<0$ (right).}
 \label{stable1}
\end{figure*}
From Fig. 1, it can be seen that for the given ranges for $w$ and $\alpha$, $\lambda^{2}$ is negative and Einstein universe is stable.

\subsection{Numerical analysis of the model}
In the following, we study numerically the effects of GUP on the dynamics of the universe. As an example, according to the allowed stability ranges for $w$, namely $-1/3 < w < 1/4$, we consider
a relativistic matter with an equation-of-state parameter $w = -0.2$. Using these equation of
state parameters in the equation (\ref{aa}), we obtain
\begin{eqnarray}
\ddot{a}+2.4 \alpha  \rho ^2 a^5-19.2 \alpha  \rho  a^3-36 \alpha  a+0.06 \rho  a=0.
\end{eqnarray}

From the above equation the corresponding scale factor of Einstein static solution is given by $a^{2}_{_{Es}} = \frac{1}{10\alpha}$.
Obviously, the phase space trajectories which are beginning precisely on the Einstein static fixed point, remain at this point indeterminately. From another point of view, trajectories which are creating in the vicinity
of this point would oscillate indefinitely near this solution.
An example of such a universe trajectory using initial conditions given by $a(0) = 1$ and $\dot{a}(0) = 0$,
with $\alpha = 0.8$ has been plotted in Fig. 2.
\begin{figure*}[ht]
 \centering
\includegraphics[width=2.5in]{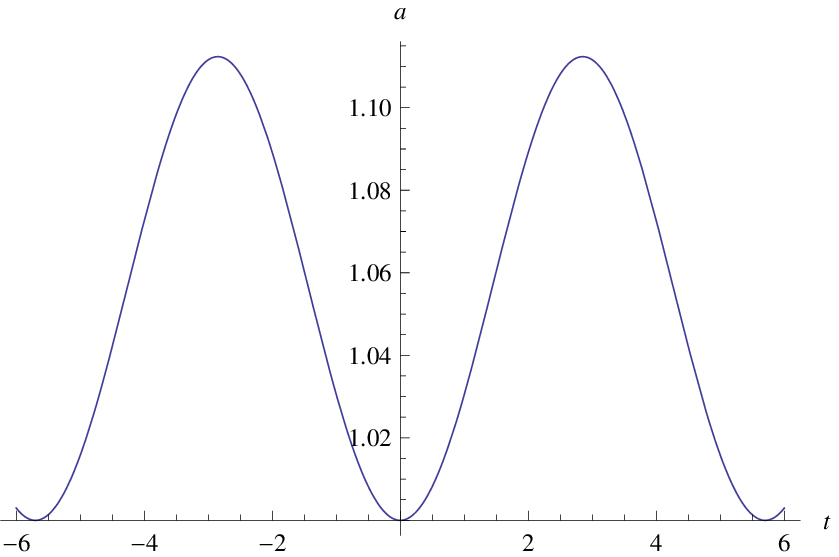}~~~~~~~~~~~~~~~
   \includegraphics[width=2in]{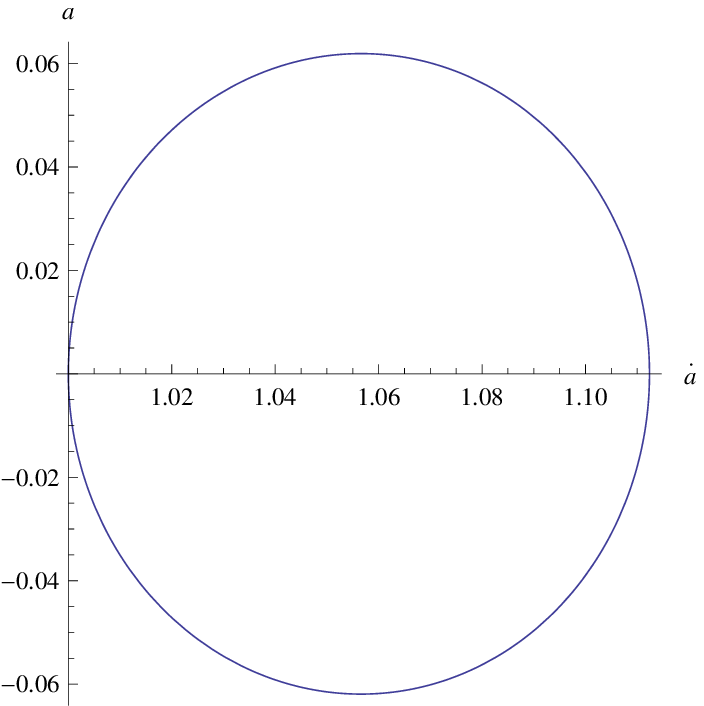}
  \caption{The evolutionary curve of the scale factor with time (left) and
the phase diagram  in space ($a$, $\dot{a}$) (right) for $w = -0.2$.}
 \label{stable}
\end{figure*}
Another example, for $\alpha = 2$ and $w = -0.3$ has been plotted in Fig. 3.
\begin{figure*}[ht]
 \centering
\includegraphics[width=2.5in]{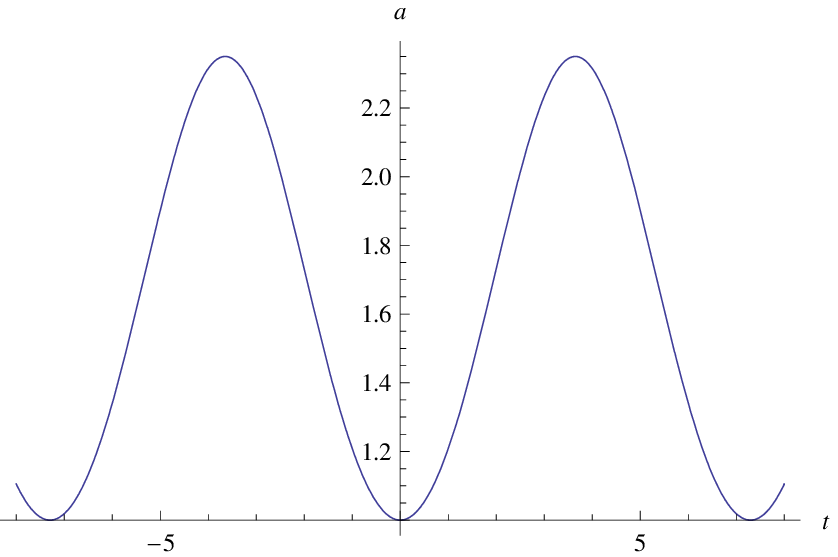}~~~~~~~~~~~~~~~
   \includegraphics[width=2.5in]{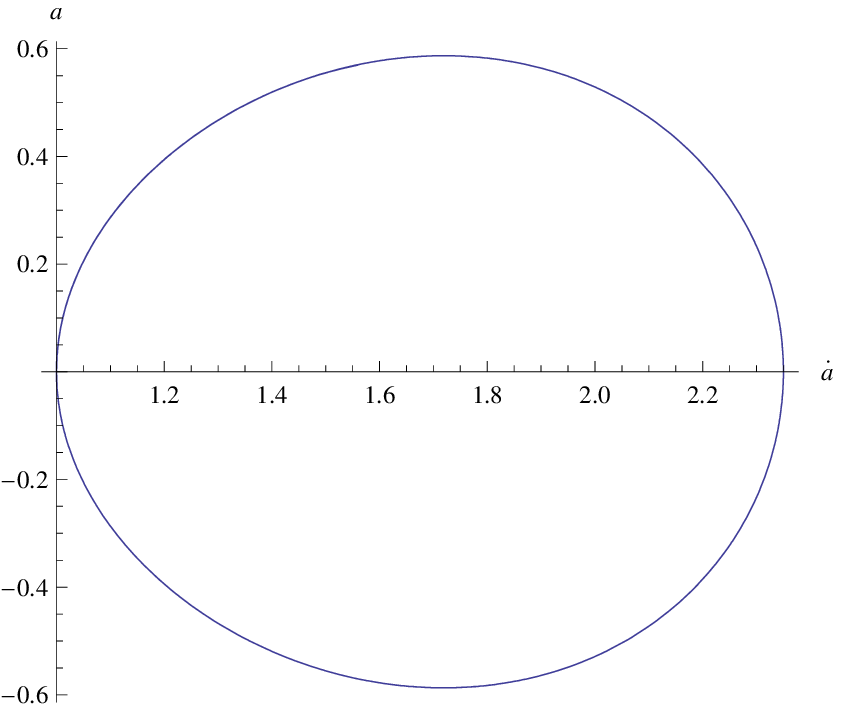}
  \caption{The evolutionary curve of the scale factor with time (left) and
the phase diagram  in space ($a$, $\dot{a}$) (right) for $w = -0.3$.}
 \label{stable2}
\end{figure*}

Note that by choosing another Equation-of-State parameter $w$ from the stability range, {\it i.e.} $-1/3<w<1/4$, one can solve numerically the equation (\ref{aa}).
\section{Inhomogeneous perturbations}
\subsection{Density perturbations}

First, we study inhomogeneous density perturbations for the simple one-component fluid models under the GUP effects.
The density perturbations in the context of FLRW universe by using $1+3$-covariant gauge-invariant approach, are characterized by $\Delta= a^{2}D^{2}\rho/\rho$, where $D^{2}$ is the covariant spatial Laplacian.
The dynamical equation of $\Delta$ for the closed FLRW universe ($K=1$), is given by \cite{Barrow,Bruni}
\begin{eqnarray}\label{dd}
&&\ddot{\Delta}+(2-6w+3c_{s}^{2})H\dot{\Delta}+\\\nonumber&&\left[\frac{12(w-c_{s}^{2})}{a^{2}}+4\pi G(3w^{2}+6c_{s}^{2}-8w-1)\rho\right]\Delta-c_{s}^{2}D^{2}\Delta-w\left(D^{2}+\frac{3}{a^{2}}\right){\cal E}=0,
\end{eqnarray}
where $c_s^2=dp/d\rho$ and ${\cal E}=(a^{2}D^{2}p-\rho c_{s}^{2}\Delta)/p$ are sound speed and the entropy perturbation for a one-component source, respectively.  For the Einstein static background model, ${\cal E} = 0$ and equation (\ref{uu}) we can rewrite equation (\ref{dd}) as
\begin{eqnarray}\label{ddd}
\\\nonumber\ddot{\Delta}_{k}+\Theta{\Delta}_{k}=0,
\end{eqnarray}
where $k$ denotes for comoving index ($D^2\rightarrow -k^2/a^2_{_{Es}}$) and $\Theta$ is given by
\begin{eqnarray}\label{dddd}
\Theta=-\frac{9 (3 w+1) \alpha  \left(279 w^2-72 w+\left(8 (3 w-1) k^2-288 w+78\right) c_s^2+3\right)}{16 (1-3
   w)^2}.
\end{eqnarray}
Equation (\ref{ddd}) shows that neutral stable against adiabatic density perturbations of the fluid for all
allowed inhomogeneous modes is generally available, except for those values of parameters $\alpha$ and $w$ for which the $\Theta$ becomes negative.

To consider the stability and instability of the Einstein static universe against adiabatic density perturbations, we obtain the following range of $w$, $\alpha$ and $k$  for the case $c_s\in\Re$ with the requirement $\Theta>0$ and $\Theta<0$ in Table 2 and Table 3, respectively.
\vspace{5mm}
\begin{center}
{\scriptsize{ Table 2: }}\hspace{-2mm} {\scriptsize Allowed ranges for $w$, $\alpha$, $k$ and $c_s$ (stable case).}\\
    \begin{tabular}{|l| l |l |l|  p{800mm} }
    \hline
   {\footnotesize$~~~~~~~~~~~~~~~~~~~w$ }& ~~{\footnotesize~~~~~~~~~~~~~~~ $c_s$ }&~~ {\footnotesize$~~~~~~~~~~~~~~~~~~~~~~~~~~~~~~~~k$ }& ~~{\footnotesize$~~\alpha$ } \\ \hline
{\footnotesize ~$\frac{-39+4k^{2}}{-144+12k^{2}}<w<\frac{1}{93}(12-\sqrt{51})  $} & ~~{\footnotesize $c_s>\sqrt{\frac{-279 w^2+72 w-3}{24 w k^2-8 k^2-288 w+78}}$}&~ {\footnotesize ~$\frac{1}{2} \sqrt{\frac{1}{10} \left(309+9 \sqrt{51}\right)}<k\leq \frac{\sqrt{\frac{87}{2}}}{2}$}& ~~{\footnotesize$\alpha>0$ } \\\hline
 {\footnotesize ~~~~~~~~~$w=\frac{1}{93}(12-\sqrt{51})  $} & ~~{\footnotesize $c_s>0$}&~ {\footnotesize ~$\frac{1}{2} \sqrt{\frac{1}{10} \left(309+9 \sqrt{51}\right)}<k\leq \frac{\sqrt{\frac{87}{2}}}{2}$}& ~~{\footnotesize$\alpha>0$ }
\\ \hline
{\footnotesize ~~~~~$-\frac{1}{3}<w<\frac{1}{93}(12-\sqrt{51})  $} & ~~{\footnotesize $c_s>0$}&~ {\footnotesize ~$\frac{1}{2} \sqrt{\frac{1}{10} \left(309-9 \sqrt{51}\right)}<k<\frac{1}{2} \sqrt{\frac{1}{10} \left(309+9 \sqrt{51}\right)}$}& ~~{\footnotesize$\alpha<0$ } \\\hline
 {\footnotesize ~$\frac{-39+4k^{2}}{-144+12k^{2}}<w<\frac{1}{93}(12-\sqrt{51})   $} & ~~{\footnotesize $c_s>\sqrt{\frac{-279 w^2+72 w-3}{24 w k^2-8 k^2-288 w+78}}$}&~ {\footnotesize ~$\frac{1}{2} \sqrt{\frac{1}{10} \left(309-9 \sqrt{51}\right)}<k<\frac{1}{2} \sqrt{\frac{1}{10} \left(309+9 \sqrt{51}\right)}$}& ~~{\footnotesize$\alpha<0$ }
\\ \hline
    \end{tabular}
\end{center}
\vspace{2mm}
\begin{center}
{\scriptsize{ Table 3: }}\hspace{-2mm} {\scriptsize Allowed ranges for $w$, $\alpha$, $k$ and $c_s$ (unstable case).}\\
    \begin{tabular}{|l| l |l |l|  p{800mm} }
    \hline
   {\footnotesize$~~~~~~~~~~~~~~~~~~~w$ }& ~~{\footnotesize~~~~~~~~~~~~~~~ $c_s$ }&~~ {\footnotesize$~~~~~~~~~~~~~~~~~~~~~~~~~~~~~~~~~k$ }& ~~{\footnotesize$~~\alpha$ } \\ \hline
{\footnotesize ~~~~~~~~$\frac{-39+4k^{2}}{-144+12k^{2}}<w<\frac{1}{3} $} & ~~{\footnotesize $c_s>\sqrt{\frac{-279 w^2+72 w-3}{24 w k^2-8 k^2-288 w+78}}$}&~ {\footnotesize $-\frac{1}{2} \sqrt{\frac{1}{10} \left(309-9
   \sqrt{51}\right)}<k<\frac{1}{2} \sqrt{\frac{1}{10} \left(309-9
   \sqrt{51}\right)}$}& ~~{\footnotesize$\alpha<0$ } \\\hline
 {\footnotesize ~~~~~~~~~~$w=\frac{1}{93}(12+\sqrt{51})  $} & ~~{\footnotesize $c_s>0$}&~ {\footnotesize ~~~~~~~~~~~~~~~~~~~~~~~$\frac{\sqrt{\frac{87}{2}}}{2}<k<2 \sqrt{3}$}& ~~{\footnotesize$\alpha<0$ }
\\ \hline
{\footnotesize ~~~~~~$-\frac{1}{3}<w<\frac{1}{93}(12-\sqrt{51})  $} & ~~{\footnotesize $c_s>0$}&~ {\footnotesize ~$\frac{1}{2} \sqrt{\frac{1}{10} \left(309-9 \sqrt{51}\right)}<k<\frac{1}{2} \sqrt{\frac{1}{10} \left(309+9 \sqrt{51}\right)}$}& ~~{\footnotesize$\alpha>0$ } \\\hline
 {\footnotesize ~$\frac{1}{93}(12-\sqrt{51})<w<\frac{-39+4k^{2}}{-144+12k^{2}} $} & ~~{\footnotesize $c_s>\sqrt{\frac{-279 w^2+72 w-3}{24 w k^2-8 k^2-288 w+78}}$}&~ {\footnotesize ~$\frac{1}{2} \sqrt{\frac{1}{10} \left(309-9 \sqrt{51}\right)}<k<\frac{1}{2} \sqrt{\frac{1}{10} \left(309+9 \sqrt{51}\right)}$}& ~~{\footnotesize$\alpha>0$ }
\\ \hline
    \end{tabular}
\end{center}

From Table 2, it can be seen that for the given ranges for $w$, $\alpha$, $k$ and $c_s$, $\Theta$ is positive and Einstein universe is stable against adiabatic density perturbations. Also, from Table 3 it can be seen that for the given ranges for $w$, $\alpha$, $k$ and $c_s$, $\Theta$ is negative and Einstein universe is unstable against adiabatic density perturbations.
It is worth mentioning that there are variety of other ranges for which one
can show the stability or instability against adiabatic density perturbations,
however we have confined ourselves to some typical ranges in the above tables.

 \subsection{Vector and tensor perturbations}

In the an isotropic and homogeneous FLRW universe, the vector perturbations of a perfect fluid are given by the comoving dimensionless vorticity defined as ${\varpi}_{a}=a{\varpi}$, with modes that are satisfying the following propagation equation \cite{Barrow,Bruni}
\begin{eqnarray}\label{37}
\dot{\varpi}_{k}+H(1-3c_{s}^2){\varpi}_{k}=0,
\end{eqnarray}
where $H$ is the Hubble parameter.
By imposing  the Einstein static universe condition, {\it i.e.} $H=0$, equation (\ref{37}) can be written as
\begin{eqnarray}\label{338}
\dot{\varpi}_k=0.
\end{eqnarray}
From equation (\ref{338}) it can be seen that in the Einstein static universe regime, the initial vector perturbations are frozen and thus for all equations of state on all scales the neutral stability against vector perturbations exists.

Next step to consider the inhomogeneous perturbations, is the tensor perturbations, namely gravitational-wave perturbations, of a perfect
fluid with density $\rho$ and pressure $p=w\rho $ that it is defined by the comoving dimensionless transverse-traceless shear $\Sigma_{a b}=a\sigma_{a b}$, with modes that are satisfying the following equation \cite{Dunsby}
\begin{eqnarray}\label{39}
\ddot\Sigma_{k}+3H\dot\Sigma_{k}+\left[\frac{k^2}{a^2}+\frac{2}{a^2}-\frac{(1+3w)\rho}{3}\right]\Sigma_{k}=0.
\end{eqnarray}
In the Einstein static universe regime, equation (\ref{39}) reads
\begin{eqnarray}\label{40}
\ddot\Sigma_{k}+\Upsilon\Sigma_{k}=0,
\end{eqnarray}
where $\Upsilon$ is given by
\begin{eqnarray}\label{400}
\Upsilon=-\left[\frac{9 (3 w+1) \left(48 (3 w-1) k^2+27 w-7\right) \alpha }{8 (1-3 w)^2}\right].
\end{eqnarray}
To obtain the above equation, we have inserted equation (\ref{uu}) in equation (\ref{39}).
This equation specifies that the neutral stability for tensor perturbations
is generally available, except for those values of parameters $\alpha$ and $w$ for which the $\Upsilon$ becomes negative.

To study the stability and instability of the tensor perturbations in the context of Einstein static universe, we obtain the following range of $w$ and $\alpha$  for the case $k\in\Re$
with the requirement $\Upsilon>0$ and $\Upsilon<0$ in Table 4 and Table 5, respectively.
\vspace{5mm}
\begin{center}
{\scriptsize{ Table 4: }}\hspace{-2mm} {\scriptsize Allowed ranges for $w$ and $\alpha$ (stable case).}\\
    \begin{tabular}{|l| l |l |  p{900mm} }
    \hline
   {\footnotesize$~~~~~~~~~~~~~~~~~~~~~~~~~~~~~~~~~w$ }& ~~{\footnotesize~ $\alpha$ }  \\ \hline
{\footnotesize ~~~~~~~~~~~~~~~~~~~~~~~~$-\frac{1}{3} < w<\frac{7+48k^{2}}{27+144k^{2}} $} & ~~{\footnotesize $\alpha>0$} \\\hline
 {\footnotesize ~~~$-1/3 > w ~~ {\rm or}~~ w > 1/3 ~~ {\rm or}~~  \frac{1}{3} > w>\frac{7+48k^{2}}{27+144k^{2}}$} & ~{\footnotesize  ~$\alpha<0$}
\\ \hline
    \end{tabular}
\end{center}
\begin{center}
{\scriptsize{ Table 5: }}\hspace{-2mm} {\scriptsize Allowed ranges for $w$ and $\alpha$ (unstable case).}\\
    \begin{tabular}{|l| l |l |  p{800mm} }
    \hline
   {\footnotesize$~~~~~~~~~~~~~~~~~~~~~~~~~~~~~~~~w$ }& ~~{\footnotesize~ $\alpha$ }  \\ \hline
{\footnotesize ~~~~~~~~~~~~~~~~~~~~~~~$-\frac{1}{3} < w<\frac{7+48k^{2}}{27+144k^{2}} $} & ~~{\footnotesize $\alpha<0$} \\\hline
 {\footnotesize ~~~$-1/3 > w ~~ {\rm or}~~ w > 1/3 ~~ {\rm or}~~  \frac{1}{3} > w>\frac{7+48k^{2}}{27+144k^{2}}$} & ~{\footnotesize  ~$\alpha>0$}
\\ \hline
    \end{tabular}
\end{center}

From Table 4, it can be seen that for the given ranges for $w$ and $\alpha$, $\Upsilon$ is positive and Einstein universe is stable against tensor perturbations. Also, from Table 5 it can be seen that for the given ranges for $w$ and $\alpha$, $\Upsilon$ is negative and Einstein universe is unstable against tensor perturbations.

\section{Conclusion}
We have discussed the existence and stability of the Einstein static universe with a minimal length in the context of GUP effects. We have shown that the radius of Einstein universe is inversely proportional to the $\alpha$. Also, we have determined the allowed intervals for the equation of state parameter such that the Einstein universe is stable, while it is dynamically belonging to a center equilibrium point. Also, we have studied the presented model under the inhomogeneous perturbations in which by fixing values of parameters $\alpha$ and $w$ stability for density, vector and tensor perturbations are generally available.
The motivation study of such a solution is the result of its essential role in
the construction of non-singular emergent oscillatory models which are past eternal, and hence can resolve
the singularity problem in the standard cosmological scenario.

\section*{Acknowledgments}

When this work was completed and ready for submission to arxiv, we noticed the appearance of a new paper
in the arxiv \cite{kh}, relevant to our paper. After a careful study, we realized that  our modified Friedmann equation (\ref{modfri}) is different from the one (7) used in \cite{kh}. In the paper \cite{kh}, the modified Friedmann equation was derived using the form of
 generalized uncertainty principle (4) with a correction term linear in the
momentum, whereas our modified Friedmann equation was derived using the different
form of  generalized uncertainty principle (\ref{ap}), with a correction term quadratic in the momentum, deduced from the Snyder non-commutative space.
Therefore, different and independent results have been obtained in these
two papers.

This work has been supported financially by a grant number 217/D/17739 from Azarbaijan Shahid Madani
University.

\end{document}